\documentstyle[twoside,fleqn,espcrc2,epsf]{article}
\newcommand{\AmS}{{\protect\the\textfont2
  A\kern-.1667em\lower.5ex\hbox{M}\kern-.125emS}}

\title{
\vspace*{-35pt}
{\normalsize \hfill {\sf UTCCP-P-94}} \\
\vspace*{-6pt}
{\normalsize \hfill {\sf Oct.\ 2000}} \\
Full QCD Light Hadron Spectrum and Quark Masses:\\
Final Results from CP-PACS\thanks{talk presented by R.~Burkhalter}}
 
\author{
{CP-PACS Collaboration}:
A.~Ali~Khan%
\address{Center for Computational Physics,
University of Tsukuba, Tsukuba, Ibaraki 305-8577, Japan \\ 
$^{\rm b}$Institute of Physics,
University of Tsukuba, Tsukuba, Ibaraki 305-8571, Japan \\
$^{\rm c}$Institute for Cosmic Ray Research, 
University of Tokyo, Kashiwa 277-8582, Japan \\
$^{\rm d}$High Energy Accelerator Research Organization (KEK), 
Tsukuba, Ibaraki 305-0801, Japan}, 
S.~Aoki$^{\rm b}$,
G.~Boyd$^{\rm a}$,
R.~Burkhalter$^{\rm a,b}$, 
S.~Ejiri$^{\rm a}$, 
M.~Fukugita$^{\rm c}$,
S.~Hashimoto$^{\rm d}$, 
N.~Ishizuka$^{\rm a,b}$,
Y.~Iwasaki$^{\rm a,b}$, 
K.~Kanaya$^{\rm a,b}$, 
T.~Kaneko$^{\rm d}$, 
Y.~Kuramashi$^{\rm d}$,
T.~Manke$^{\rm a}$\thanks{address after 1 Feb., 2000:
        Department of Physics, Columbia University,
        538 W 120th St., New York, NY 10027, USA},
K.~Nagai$^{\rm a}$,
M.~Okawa$^{\rm d}$, 
H.P.~Shanahan$^{\rm a}$\thanks{address after 15 Sept., 2000:
        Department of Biochemistry and Molecular
        Biology, University College London, London, England, UK},
A.~Ukawa$^{\rm a,b}$ and T.~Yoshi\'e$^{\rm a,b}$
}

\begin{document}

\begin{abstract}
We present the final results of the CP-PACS calculation of the light hadron
spectrum and quark masses with two flavors of dynamical quarks.
Simulations are made with a renormalization-group improved gauge action and
a mean-field improved clover quark action for sea quark masses
corresponding to $m_{\rm PS}/m_{\rm V} \approx 0.8$--0.6 and the lattice
spacing $a=0.22$--0.11 fm.  For the meson spectrum in the continuum limit 
a clearly improved agreement with experiment is observed compared to the
quenched case, demonstrating the importance of sea quark 
effects.  For light quark masses we obtain 
$m_{ud}^{\overline{MS}}(2GeV)=3.44^{+0.14}_{-0.22}$~MeV and 
$m_s^{\overline{MS}}(2GeV)=88^{+4}_{-6}$~MeV ($K$-input) and
$m_s^{\overline{MS}}(2GeV)=90^{+5}_{-11}$~MeV ($\phi$-input), 
which are reduced by about 25\% compared to the values in quenched QCD.
\end{abstract}

\maketitle

\section{INTRODUCTION}

Recent progress in lattice calculations of the quenched light hadron
spectrum~\cite{cp-pacs} has demonstrated clearly that effects of dynamical 
sea quarks have to be included for a precision comparison with 
the experimental spectrum.
As a first step towards the real world we have pursued a systematic
calculation in two-flavor QCD over the last 3 years~\cite{lat98,lat99}. 
In this article we present a summary of 
the final results on the light hadron spectrum and 
quark masses from this calculation.  

In order to deal with the increased demand in computer time for full QCD we
have used an RG-improved gauge action and a tadpole-improved SW clover
quark action, which allows us to simulate at coarse lattices in the range
$a=0.22$--0.11 fm~\cite{lat98,lat99}. 
Since Lattice'99~\cite{lat99} we have doubled the statistics at the smallest 
lattice spacing on a $24^3\times 48$ lattice, with which we have reached 
our planned target of three lattice spacings, each with a similar amount of 
statistics at each simulated sea quark mass. 
The final parameters of our simulations are summarized in
Table~\ref{tab:param}.

We have also made a set of
simulations in quenched QCD using the same improved actions in the same
range of lattice spacings and with a similar range of quark masses. This
allows us to compare in detail spectrum results in full and quenched QCD
and to make an independent confirmation of the findings in quenched QCD in
Ref.~\cite{cp-pacs}. Parameters of quenched simulations are summarized in
Table~\ref{tab:quench}.

\begin{table*}[tb]
\setlength{\tabcolsep}{0.40pc}
\begin{center}
\caption{\label{tab:param}
Parameters and final statistics of two-flavor QCD
simulations.} 
\begin{tabular}{ccccccccc}
\hline
$\beta$ & $L^3\times T$ & $c_{SW}$ & $a$[fm] & $La$[fm] 
& \multicolumn{4}{c}
  {$m_{\rm PS}/m_{\rm V}$ for sea quarks : \# HMC trajectories} \\
\hline
1.80 & $12^3{\times}24$ & $1.60$ & 0.215(2) & 2.58(3) 
        & 0.807(1):6250  & 0.753(1):5000 & 0.694(2):7000 & 0.547(4):5250 \\ 
1.95 & $16^3{\times}32$ & $1.53$ & 0.155(2) & 2.48(3) 
        & 0.804(1):7000 & 0.752(1):7000 & 0.690(1):7000 & 0.582(3):5000 \\ 
2.10 & $24^3{\times}48$ & $1.47$ & 0.108(1) & 2.58(3) 
        & 0.806(1):4000 & 0.755(2):4000 & 0.691(3):4000 & 0.576(3):4000 \\ 
\hline
\end{tabular}
\end{center}
\vspace{-6mm}
\end{table*}

\section{SEA QUARK EFFECTS IN THE LIGHT HADRON SPECTRUM}

A novel feature in the spectrum analysis with two-flavor QCD 
is the dependence of hadron masses on both the sea and the valence
quark mass. We take this into account by a combined quadratic
ansatz.  For vector mesons and decuplet baryons the formula has the form,
\begin{eqnarray}
m_{V,D} & = & A + B_s \mu_{sea} + B_v \mu_{val} \nonumber \\
        &   &  +  C_s \mu_{sea}^2 + C_v \mu_{val}^2 + 
              C_{sv} \mu_{sea} \mu_{val},    
\label{eq:vec-dec-fit}
\end{eqnarray}
where $\mu_{val}$ represents the mass squared $m_{PS}^2$ of a pseudo scalar
meson made of a valence quark-antiquark pair and $\mu_{sea}$ stands for a
$PS$ meson made of sea quarks.  Since octet baryon masses are not functions
of the average valence quark mass we use a more complicated formula with 12
parameters. For quenched results with improved actions terms in
$\mu_{sea}$ are dropped and we find that vector meson and decuplet baryon
masses are well described by a linear ansatz in $\mu_{val}$.
Physical masses in full QCD are obtained from Eq.~\ref{eq:vec-dec-fit} at
$\mu_{sea}=\mu_{ud}$ and $\mu_{val}=\mu_{ud}$ or $\mu_s$, which are fixed 
using $\pi$, $\rho$ and either $K$ or $\phi$ meson masses as input. 

\begin{table}[b]
\setlength{\tabcolsep}{0.19pc}
\vspace{-12mm}
\begin{center}
\caption{\label{tab:quench}
Parameters of quenched QCD simulations.} 
\begin{tabular}{lllllll}
\hline
\multicolumn{3}{l}{$16^3{\times}32$} &  &  
\multicolumn{3}{l}{$24^3{\times}48$}
\\
 \cline{1-3} \cline{5-7}
 $\beta$ &   $a$~[fm]   & $La$~[fm]  & &  $\beta$  &  $a$~[fm] & $La$~[fm]  \\
 \cline{1-3} \cline{5-7}
 2.187 & 0.200(2) & 3.21(3) & &  2.416 & 0.145(2) & 3.47(4)  \\  
 2.214 & 0.190(2) & 3.05(3) & &  2.456 & 0.133(1) & 3.19(3) \\
 2.247 & 0.181(2) & 2.89(3) & &  2.487 & 0.128(1) & 3.08(3)  \\
 2.281 & 0.177(2) & 2.82(3) & &  2.528 & 0.121(1) & 2.89(3)  \\
 2.334 & 0.163(2) & 2.61(3) & &  2.575 & 0.113(1) & 2.71(3)  \\
 \hline
\end{tabular}
\end{center}
\end{table}

Since our simulations with improved action employ a mean-field clover 
coefficient $c_{SW}=P^{-3/4}$ for the quark action, the leading scaling 
violation is $O(g^2a)$. Here
$g^2$ is the renormalized coupling $g^2_{\overline{MS}}(\mu)$~\cite{alpha},
evaluated at a fixed scale $\mu$.  Higher order contributions consist
of terms of $O(g^4 a \log(a))$, $O(a^2)$ etc.  However, the number and range
of lattice spacings restrict our choice of fitting form to the leading
scaling violation term, linear in $a$, which we apply for
extrapolations to the continuum limit both in quenched and full QCD.

\begin{figure}[h]
\vspace{-6mm}
\centerline{ \epsfxsize=6.4cm \epsfbox{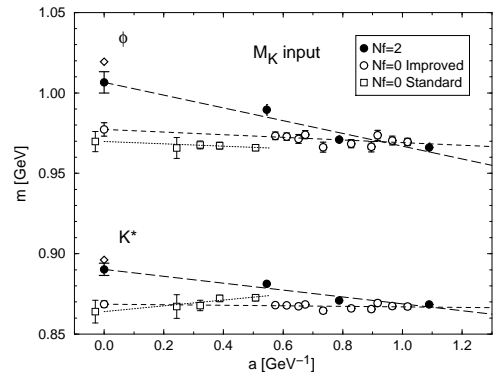} }
\vspace{-11mm}
\caption{Meson masses in two-flavor (filled
symbols) and quenched (open symbols) QCD.}
\label{fig:ContMes}
\vspace{-7mm}
\end{figure}

Our main results for the meson spectrum are given in Fig.~\ref{fig:ContMes}
where $\phi$ and $K^*$ meson masses obtained with $K$-input are plotted. 
The new quenched results with the improved action 
nicely confirm the finding with the standard action
in Ref.~\cite{cp-pacs} that the quenched QCD prediction for the $\phi$ 
and $K^*$ masses is smaller by about 4\% than experiment.  
On the other hand, the two-flavor results exhibit a larger
slope than those of quenched QCD, and the values linearly extrapolated 
to the continuum limit lie within 1\% of the experimental values. 

We find this result to be a clear evidence of the importance of sea 
quark effects.  The observed trend that the meson hyperfine splitting 
is enlarged by such effects is consistent with a qualitative view that 
the spin-spin coupling in quenched QCD is reduced compared to full QCD 
due to a faster running of the quenched QCD coupling constant with 
scale.

\begin{figure}[htb]
\centerline{ \epsfxsize=6.4cm \epsfbox{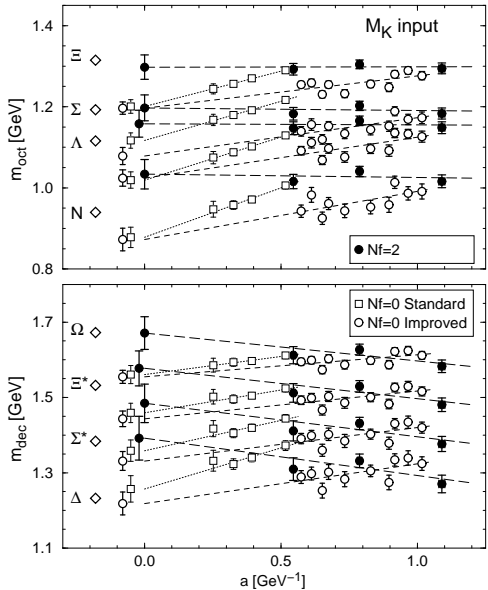} }
\vspace{-11mm}
\caption{Corresponding plot for baryon masses.}
\label{fig:ContBar}
\vspace{-7mm}
\end{figure}

Our baryon mass results are collected in Fig.~\ref{fig:ContBar}.  Starting
with quenched QCD, our new results with the improved action 
show sizable statistical fluctuations.  With ten points of data,
however, one observes a trend that the magnitude of scaling violation is
smaller than for the standard action.  Linearly
extrapolating to the continuum limit, the spectrum for octet and
decuplet baryons is consistent between the two actions.

A worry with the baryon spectrum is possible finite-size effects.  
As we see from Table~\ref{tab:quench}, however, the spatial size 
mostly lies in the range $La\approx 2.8$-3.5~fm, and only in two cases
goes down to $La\approx 2.7-2.6$~fm. Finite-size effects are suppressed in 
quenched QCD due to $Z(3)$ symmetry\cite{z3}.  Past studies\cite{fss} also 
indicate that $La\approx 3$~fm is sufficient to contain the magnitude of 
finite-size effects for baryons within a few \% level.  We therefore consider 
that our quenched baryon results do no suffer seriously from finite-size 
contaminations. 

Clear conclusions are difficult to draw for two-flavor results.
Since quenched results are generally smaller than experiment, 
we expect sea quark effects to push up the baryon masses.  While this 
is a general trend observed in our data, the nucleon and the $\Delta$ 
masses turn out higher than experiment, whereas a better agreement is 
seen for heavier strange baryons. 

Finite-size effects are severer in full QCD with dynamical sea quarks 
since suppression due to $Z(3)$ symmetry is no longer effective. 
It is probable that the spatial size of 2.5~fm in full QCD, which is
smaller than the smallest quenched lattice, is too small to avoid 
finite-size effects for baryons, especially those made of light up and 
down quarks.  Further studies on larger spatial sizes are needed to 
convincingly demonstrate sea quark effects in the baryon mass spectrum. 

\section{LIGHT QUARK MASSES}

The light quark masses are fundamental parameters of Nature which 
can only be determined by hadron spectrum calculations. 
In Table~\ref{tab:final} we present our two-flavor results for up-down 
and strange quark masses in the $\overline{MS}$ scheme at $\mu=2$~GeV, 
and compare them with quenched results. These results have been published 
in Ref.~\cite{quarkletter}. 
The quenched results, which are consistent between the two types of 
actions, overestimate the quark mass by about 25\% compared to 
two-flavor QCD.  Furthermore, a discrepancy in the strange quark mass in 
quenched QCD depending on the experimental input is reduced to a level 
well contained in the error of about 10\%.  Both these features demonstrate
a significant improvement in the determination of light quark masses through 
inclusion of two flavors of dynamical quarks.  

\newcommand{\STRUT}{\rule{0in}{2.5ex}}
\begin{table}[tb]
\setlength{\tabcolsep}{0.25pc}
\caption{\label{tab:final}
Final results for quark masses in $\overline{MS}$ scheme at
$\mu=2$~GeV (in MeV).}
\begin{tabular}{ccccc}
\hline
        & action & $m_{ud}$  & \multicolumn{2}{c}{$m_s$} \\
        &        &           & $K$-input & $\phi$-input \\
\hline
$N_f=2$\STRUT & impr.  & $3.44^{+0.14}_{-0.22}$  & 
            $88^{+4}_{-6}$  &  $90^{+5}_{-11}$ \\[0.5ex]
$N_f=0$ & impr.  & $4.36^{+0.14}_{-0.17}$  &
                   $110^{+3}_{-4}$  &  $132^{+4}_{-6}$ \\[0.5ex]
$N_f=0$ & stand.\cite{cp-pacs} & 4.57(18)                &
                   $116(3)$        &  $144(6)$ \\[0.5ex]
\hline
\end{tabular}
\vspace{-7mm}
\end{table}

\begin{figure}[htb]
\centerline{ \epsfxsize=6.8cm \epsfbox{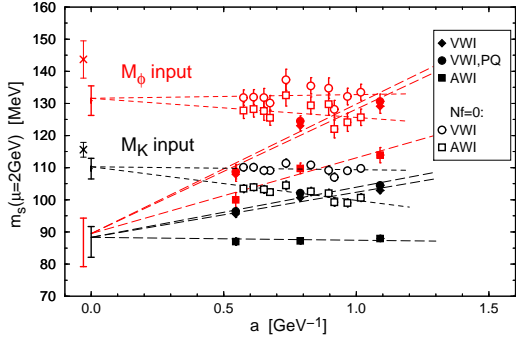} }
\vspace{-12mm}
\caption{Continuum extrapolation of strange quark mass for improved
actions. Results from standard action in Ref.\cite{cp-pacs} are also shown 
(crosses).}
\label{fig:ms}
\vspace{-7mm}
\end{figure}

Our results are obtained from analyses of two definitions of quark
mass, one derived from the axial Ward identity $\nabla_{\mu}A_{\mu}(x) =
2am_q^{\rm AWI}P(x)$, and another $m_q^{\rm VWI} = (1/\kappa -
1/\kappa_c)/2a$ derived from the vector Ward identity. For the latter, two
choices for $\kappa_c$, either the one in full QCD or in partially quenched
chiral limit, are examined.  In Fig.~\ref{fig:ms} we illustrate the continuum
extrapolation for the strange quark mass. Lines show a linear fit assuming a
common value in the continuum limit, which we take to be our central value.

A crucial point is the estimation of systematic 
errors, which arise from (i) chiral extrapolation, (ii) renormalization 
factor, and (iii) continuum extrapolation. 
We estimate the error from (iii) by making linear fits separately to each 
definition of quark mass and taking the spread of results. 
The error due to chiral extrapolation (ii) is estimated by replacing
quadratic terms in Eq.~\ref{eq:vec-dec-fit} by terms with power $3/2$ which
also appear in chiral perturbation theory and by introducing cubic terms in
fits of pseudo scalar mesons. 

\begin{table}[b]
\setlength{\tabcolsep}{0.18pc}
\vspace{-6mm}
\caption{\label{tab:errors}
Contributions to total error in continuum limit.}
\begin{tabular}{ccccc}
\hline
                     & stat.    & chiral   & Z-factor & cont.ext.\\
\hline
$m_{ud}$             & $+2.6\%$ & $+1.2\%$ & $+2.3\%$ & $+1.7\%$  \\
                     & $-2.6\%$ & $-2.3\%$ & $-5.0\%$ & $-2.3\%$  \\
$m_s$ ($K$-input)    & $+2.4\%$ & $+1.6\%$ & $+2.2\%$ & $+1.4\%$  \\
                     & $-2.4\%$ & $-2.2\%$ & $-5.6\%$ & $-2.8\%$ \\
$m_s$ ($\phi$-input) & $+4.8\%$ & $+1.5\%$ & $+1.7\%$ & $+0.9\%$  \\
                     & $-4.8\%$ & $-7.6\%$ & $-6.9\%$ & $-1.6\%$  \\
\hline
\end{tabular}
%\vspace{-6mm}
\end{table}

For the renormalization factors relating the lattice quark mass 
to the $\overline{MS}$ mass in the continuum, we employ the perturbative 
values at one loop\cite{taniguchi} given by 
$Z_m = 1 + 0.0400 g^2_{\overline{MS}}(\mu)$ for VWI quark mass and 
$Z_A/Z_P =  1 + 0.0308 g^2_{\overline{MS}}(\mu)$ for AWI quark mass. 
These expressions show that the magnitude of one-loop corrections is 
about 10\%. One loop corrections in improvement coefficients have a similar 
size. To estimate further corrections from higher order terms,  
we compare results for two definitions of the mean-field improved
coupling constant: $g^{-2}_{\overline{MS}}(1/a) =
(3.648W_{1\times 1}-2.648W_{1\times 2})\beta/6 - 0.1006 + 0.03149N_f$,  
appropriate for the RG-improved gluon action and alternatively 
$g^{-2}_{\overline{MS}}(1/a) = W_{1\times 1}\beta/6 + 0.2402 + 0.03149N_f$.
For the matching scale we use the two values $\mu=1/a$ and $\mu=\pi/a$.

The systematic errors estimated as above are listed in Table~\ref{tab:errors}.
We calculate the total error by adding them by quadrature.

\section{CONCLUSIONS}

Our first systematic study of $N_f=2$ full QCD clearly showed the existence 
of sea quark effects in the meson spectrum and quark masses. Further
progress can be expected in the near future from calculations with larger
lattices, smaller quark masses and, most importantly, inclusion
of dynamical strange quark.

This work is supported in part by Grants-in-Aid 
of~the~Ministry~of~Education~(Nos.~10640246,
10640248,~10740107,~11640250,~11640294,
11740162,~12014202,~12304011,~12640253, 12740133).  
AAK and TM are supported by JSPS Research for the Future Program
(No. JSPS-RFTF 97P01102).
SE, TK, KN and HPS are JSPS Research Fellows.


\vspace{-6pt}
 
\begin{thebibliography}{9}

\vspace{-3pt}

\bibitem{cp-pacs}
S.~Aoki {\it et al.} (CP-PACS Collaboration), 
Phys.\ Rev.\ Lett.\ {\bf 84}, 238(2000); 
Nucl.\ Phys.\ B(Proc. Suppl.){\bf 73}, 189(1999).

\bibitem{lat98}
R. Burkhalter, 
Nucl.\ Phys.\ B (Proc. Suppl.) {\bf 73}, 3(1999);
S.~Aoki {\it et al.} (CP-PACS Collaboration),
{\it ibid.} {\bf 73}, 192(1999).

\bibitem{lat99}
A.~Ali~Khan {\it et al.} (CP-PACS Collaboration), 
Nucl.\ Phys.\ B (Proc. Suppl.) {\bf 83-84}, 176(2000).

\bibitem{alpha} 
M.~L\"uscher {\it et al.}, Nucl.\ Phys.\ {\bf B384}, 168(1992); 
{\it ibid.} {\bf B389}, 247(1993). 

\bibitem{z3}
S.~Aoki {\it et al.,} Phys. Rev. D{\bf 50}, 486 (1994) and
earlier references therein.

\bibitem{fss}
See, {\it e.g.}, MILC Collaboration, C.~Bernard {\it et al.},
Nucl. Phys. {\bf B} (Proc. Suppl.) {\bf 60A}, 3 (1998).
 
\bibitem{quarkletter}
A.~Ali~Khan {\it et al.} (CP-PACS Collaboration), Phys. Rev. Lett., in
print (hep-lat/0004010).

\bibitem{taniguchi}
S.~Aoki {\it et al.,} Phys. Rev. D {\bf 58}, 074505 (1998);
Y.~Taniguchi and A.~Ukawa, {\it ibid.} {\bf 58}, 114503 (1998).


\end{thebibliography}
\end{document}